\documentclass{appolb}
\usepackage{epsfig,amsmath,cite,floatflt,amssymb}
\def\cO#1{{\CMcal O}\left( {#1} \right)}
\def\unds{{\textsc{s}}}
\def\as{\alpha_\unds}
\def\asb{{\bar \alpha}_\unds}
\def\ga{\gamma}
\def\om{\omega}
\def\ca{C_A}
\def\cf{C_F}
\def\nf{n_{\! f}}
\def\shat{{\hat s}}
\def\sh{\shat}
\def\cO#1{{\cal O}\left( {#1} \right)}
\def\cF{{\cal F}}

\def\d{\mathrm{d}}
\def\gabar{{\bar \ga}}
\def\eff{{\textrm{eff}}}
\def\l{\left}
\def\r{\right}
\def\chit{{\tilde \chi}}
\def\coll{{\textrm{coll}}}
\def\MSbar{\overline{\mbox{\scriptsize \textsc{ms}}}}
\def\hom{\frac{\om}{2}}
\def\gaga{\ga^*\ga^*}
\def\gev{\,\mbox{GeV}}
\def\cnf{{\it cf.}\comsp}

\def\ombar{{\bar \om}}

\newcommand\hepph[1]{hep-ph/#1}
\newcommand\hepth[1]{hep-th/#1}
\newcommand\hepex[1]{hep-ex/#1}

\newcommand\jhep[3]{{{\it \textsc{JHEP} }{\bf #1} (#2) #3}}

\newcommand\npb[3]{{{\it Nucl. Phys. }{\bf B #1} (#2) #3}}

\newcommand\plb[3]{{{\it Phys. Lett. }{\bf B #1} (#2) #3}}
\newcommand\pr[3]{{{\it Phys. Rev. }{\bf #1} (#2) #3}}

\newcommand\prd[3]{{{\it Phys. Rev. }{\bf D #1} (#2) #3}}

\newcommand\zpc[3]{{{\it Z. Physik }{\bf C #1} (#2) #3}}

\newcommand\sjnp[3]{{{\it Sov. J. Nucl. Phys. }{\bf #1} (#2) #3}}
\newcommand\jetp[3]{{{\it Sov. Phys. \textsc{JETP} }{\bf #1} (#2) #3}}

\newcommand\jetpl[3]{{{\it \textsc{JETP} Lett. }{\bf #1} (#2) #3}}

\def\epjc#1#2#3{{ \it Eur. Phys. J. } {\bf {C#1}} (#2) #3}

\begin{document}

\title{%
An introduction to leading and next-to-leading BFKL%
\thanks{%
  Presented at the Cracow School of Theoretical Physics, XXXIX Course, 1999.\\
  Work supported by E.U. QCDNET contract FMRX-CT98-0194.
}%
}
\author{Gavin P. Salam
\address{
\vspace{-5cm}
\begin{flushright}
   Bicocca--FT--99--35\\
   hep-ph/9910492\\
   October 1999
\end{flushright}
\vspace{2.8cm}
INFN --- Sezione di Milano, \\ Via Celoria 16, Milano 20133, Italy}
}
\maketitle
\begin{abstract}
  Of late, the field of BFKL physics has been the subject of
  significant developments. The calculation of the NLL terms was
  recently completed, and they turned out to be very large. Techniques
  have been proposed to resum these corrections. These lectures
  provide an introduction to the BFKL equation and some of the recent
  developments, using DGLAP evolution as the starting point.
\end{abstract}
  
\section{Introduction}

Some twenty-five years ago Balitsky, Fadin, Kuraev and Lipatov (BFKL)
set out to determine the high-energy behaviour of the scattering of
hadronic objects within perturbative \textsc{QCD}.  They found terms
going as $(\as\ln s)^n$, where $s$ is the squared centre-of-mass
energy. Since $\ln s$ is large it can compensate the smallness of
$\asb$ and thus it was necessary to sum this whole series of
\emph{leading logarithmic} (\textsc{LL}) terms. The result was that
the cross section should grow as a power of the squared centre-of-mass
energy $s$ \cite{BFKL}. For the values of $\as\simeq 0.2$ that are
typically relevant, this power comes out as being of the order of
$0.5$.

Over the past few years much experimental effort has been devoted
towards observing this phenomenon, and the conclusion has consistently
been that while the cross sections do rise, that rise is much slower
than $s^{0.5}$ (see for example \cite{H1omeff,ZEUSF2,L3,OPAL,FJ}).

The solution to this problem was to have been in the next-to-leading
corrections to the \textsc{BFKL} equation, terms $\as (\as \ln s)^n$,
which have been calculated over the past ten years \cite{NLL}.  The
various contributions were put together last year \cite{FL,CC98}, and
to the consternation of the community turned out to be larger than the
leading contribution, giving cross sections that were not even
positive-definite \cite{Ross,Levin}.

These lectures will illustrate the origin of some of the main features
of both the leading and next-to-leading \textsc{BFKL} equations, using
as a basis the constraints provided by the \textsc{DGLAP} equation,
and follow on with a discussion, based on \cite{GPS,CCS}, of how to
solve the problems that arise at next-to-leading order.

After a brief definition of the problem in the next subsection,
section~\ref{sec:LL} discusses the \textsc{DGLAP} equation as relevant
for high-energy scattering, and shows how it can naturally be extended
to the give the BFKL equation \cite{BFKL}. This is followed by an
illustration of the lack of agreement of the latter with experimental
data.  Section~\ref{sec:nll} derives the main features of the
next-to-leading corrections to \textsc{BFKL} and discusses some of the
problems that ensue from their inclusion.  Section~\ref{sec:resum}
looks at how one can go beyond next-to-leading order and
section~\ref{sec:outlook} concludes.

\subsection{The problem}

\begin{floatingfigure}[r]{0.4\textwidth}
  \begin{center}
    \begin{picture}(0,0)%
\epsfig{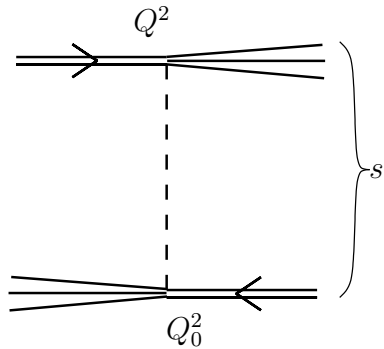}%
\end{picture}%
\setlength{\unitlength}{4144sp}%
\begingroup\makeatletter\ifx\SetFigFont\undefined%
\gdef\SetFigFont#1#2#3#4#5{%
  \reset@font\fontsize{#1}{#2pt}%
  \fontfamily{#3}\fontseries{#4}\fontshape{#5}%
  \selectfont}%
\fi\endgroup%
\begin{picture}(2195,2020)(608,-1474)
\put(1688,-1474){\makebox(0,0)[b]{\smash{\SetFigFont{12}{14.4}{\familydefault}{\mddefault}{\updefault}$Q_0^2$}}}
\put(1493,366){\makebox(0,0)[b]{\smash{\SetFigFont{12}{14.4}{\familydefault}{\mddefault}{\updefault}$Q^2$}}}
\put(2791,-511){\makebox(0,0)[lb]{\smash{\SetFigFont{12}{14.4}{\familydefault}{\mddefault}{\updefault}$s$}}}
\end{picture}

    \caption{High-energy collision of two hadronic objects.}
    \label{fig:colsn}
  \end{center}
\end{floatingfigure}
Let us first define a little more carefully the problem to be
addressed. We want to study collisions of two perturbative hadronic
objects, figure~\ref{fig:colsn}, where the squared centre-of-mass
energy $s$ is much larger than the typical transverse scales $Q^2$,
$Q_0^2$ of the two objects, which in turn are much larger than the
\textsc{QCD} scale, $\Lambda^2$, in order for the problem to be
perturbative.  This is of phenomenological relevance for certain
features of small-$x$ deep-inelastic scattering (\textsc{DIS}) at
\textsc{HERA}, high-energy $\gamma^*\gamma^*$ scattering at
\textsc{LEP} and the \textsc{NLC}, and configurations at the Tevatron
and \textsc{LHC} involving jets that are widely separated in rapidity.
It is also of theoretical interest since the large parton densities
that arise at high energies can lead to novel effects such as parton
recombination and multiple perturbative scatterings.

\section{Leading-logarithmic order}
\label{sec:LL}

\subsection{Deep inelastic scattering}
\label{sec:dis}

\begin{figure}[tb]
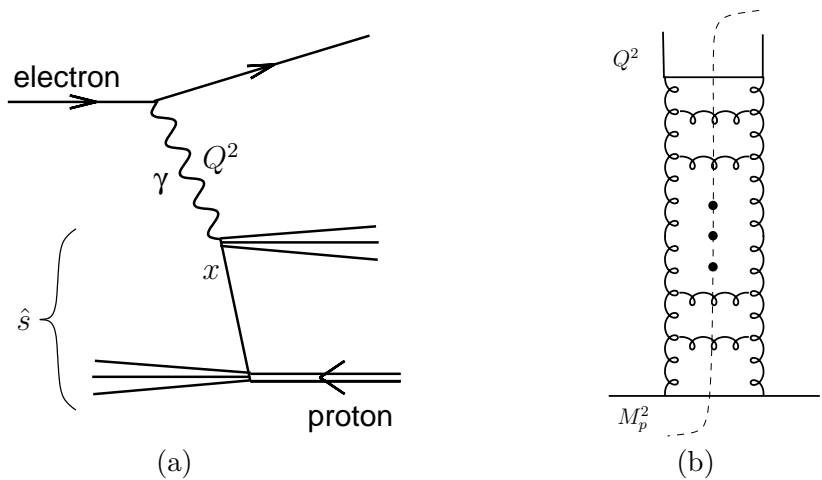

  \begin{minipage}[b]{0.45\textwidth}
    \begin{center}
      \input{dis-labelled.pstex_t}\\(a)
    \end{center}
  \end{minipage}\hfill
  \begin{minipage}[b]{0.45\textwidth}
    \begin{center}
      \resizebox{!}{0.3\textheight}{\input{ladder.pstex_t}}
      \\(b)
    \end{center}
  \end{minipage}
  \caption{(a) Deep inelastic scattering. (b) Cut ladder diagram for
    the evolution of the parton distributions.}
  \label{fig:dislad}
\end{figure}

Rather than entering straight into the problem of general high-energy
scattering, it is helpful to consider first high-energy scattering in
which one of the two hadronic objects is much smaller than the other,
\ie deep inelastic scattering, figure~\ref{fig:dislad}a. We have the
collision of a proton (of mass $M_p^2$, equivalent to $Q_0^2$ of
figure~\ref{fig:colsn}) with a photon of virtuality $Q^2 \gg M_p^2$,
which we will view as our second hadronic object. The photon-proton
squared centre-of-mass energy is $\sh$.  High-energy scattering
in this system, $\shat \gg Q^2$, is generally referred to as small-$x$
scattering because Bjorken-$x$ is $=Q^2/\shat \ll 1$.

As is well known, to correctly treat such collisions it is necessary
to resum terms $(\as \ln Q^2)^n$, because the smallness of $\as$ is
compensated by the large size of $\ln Q^2$. This is \textsc{DGLAP} 
\cite{DGLAP} or collinear resummation, or renormalisation group
evolution.

The cross section is proportional to the quark distribution at scale
$Q^2$, which is related to the quark distribution at another scale
$Q_0^2$ by
\begin{multline}
  x q(x,Q^2) = x q(x,Q_0^2) + \as \ln \frac{Q^2}{Q_0^2} \int \d z_1
  \, P_{qq}(z_1) \, \frac{x}{z_1} \,q\left(\frac{x}{z_1},Q_0^2\right)
  \\ + \as \ln \frac{Q^2}{Q_0^2} \int \d z_1
  \, P_{qg}(z_1)  \, \frac{x}{z_1} \,g\left(\frac{x}{z_1},Q_0^2\right)
  + \ldots
\end{multline}
In an appropriate gauge this can interpreted as the first in a set of
ladder diagrams (figure \ref{fig:dislad}b), whose rungs are strongly
ordered in $\ln Q^2$.

To understand the type of ladder that dominates at small $x$, we need
to look at the splitting functions. A quark ladder (with gluon rungs)
involves iteration of the $P_{qq}$ splitting function:
\begin{equation}
  P_{qq}(z) = \frac{\cf}{2\pi} \left[ \frac{1+z^2}{(1-z)_+} + \frac32
    (1-z)\right],
\end{equation}
while a gluon ladder (with gluon rungs) involves the $P_{gg}$
splitting function,
\begin{equation}
  \label{eq:pgg}
  P_{gg}(z) = \frac{\ca}{\pi} \left[ \frac1z + \frac1{(1-z)_+} - 2 +
    z(1-z)\right] + \delta(1-z) \beta_0\,.
\end{equation}
At small $z$, $P_{qq}$ is constant while $P_{gg}$ grows as $1/z$. So
at small $x$, gluon ladders with repeated iterations of
$P_{gg}(z\ll1)$ dominate, \ie we have strong ordering in $z$. 

With this is mind, let us examine the properties of the
\emph{unintegrated} gluon distribution:
\begin{equation}
  \cF(x,Q^2) = x \frac{\d g(x,Q^2)}{\d Q^2}\,,
\end{equation}
and start with a simple (though not entirely physical) initial condition
\begin{equation}
  \label{eq:Finit}
  Q^2 \cF^{(0)}(x,Q^2) = \Theta(1-x) \; \Theta\!\!\l(\frac{Q^2}{Q_0^2}
  - 1\r)\,,
\end{equation}
where the $Q^2$ factor is included on dimensional grounds ($g(x,Q^2)$
is dimensionless). Using the purely gluonic \textsc{DGLAP} equation in
differential form,
\begin{equation}
  \label{eq:DGLAPdiff}
  Q^2 x \frac{\d  g(x,Q^2)}{\d Q^2} = \as \int^1_x \,\d z P_{gg}(z)\,
  \frac{x}{z} g\l(\frac{x}{z},Q^2\r),
\end{equation}
and rewriting it in terms of the unintegrated gluon distribution, we
obtain the first-order contribution to $\cF$,
\begin{align}
  Q^2 \cF^{(1)}(x,Q^2) &= \as \int_x^1 \d z_1 P_{gg}(z_1) \int^{Q^2}
  \d k_1^2 \,\cF^{(0)}\! \left(\frac{x}{z_1},
    k_1^2\right) \nonumber\\
  &\simeq \asb \int_x^1 \frac{\d z_1}{z_1} \int^{Q^2} \d k_1^2
  \,\cF^{(0)}\!\left(\frac{x}{z_1}, k_1^2\right) = \asb \ln \frac1x \ln
  \frac{Q^2}{Q_0^2} \,,\label{eq:DLF1}
\end{align}
where $\asb=\as\ca/\pi$ has been introduced as a notational shorthand
and a factor $\Theta(Q^2-Q_0^2)$ is implicitly understood to be
contained in the result. We retain only the $1/z$ part of the
splitting function because the other parts lead to contributions
lacking the factor $\ln 1/x$ and so much smaller than \eqref{eq:DLF1}.

The second-order contribution is
\begin{multline}
  Q^2\cF^{(2)}(x,Q^2) = \asb \int_x^1 \frac{\d z_2}{z_2} \int^{Q^2}
  \d k_2^2 \, \cF^{(1)}\!\left(\frac{x}{z_2},
    k_2^2\right)
  = \frac{\asb^2}{(2!)^2} \ln^2 \frac1x \ln^2 \frac{Q^2}{Q_0^2}\,,
\end{multline}
By iteration one sees that the $\cO{\asb^{n}}$ contribution is
\begin{equation}
  \label{eq:dlseries}
  Q^2 \cF^{{(n)}}(x,Q^2) 
  = \frac{1}{(n!)^2}
  \left(\asb\ln \frac1x \ln \frac{Q^2}{Q_0^2}\right)^n
  \Theta(Q^2 - Q_0^2)\,. 
\end{equation}
Since every power of $\asb$ is accompanied by two logarithms, this is
referred to as a double-logarithmic (\textsc{DL}) series. It resums ladders in
which there is strong ordering of both the transverse and longitudinal
momenta along the ladders: $k_n^2/k_{n-1}^2 \gg 1$ and $z_n \ll 1$
respectively.

\subsection{Summing the \textsc{DL} series}

\begin{figure}[htbp]
  \begin{center}
    \resizebox{1.0\textwidth}{!}{\input{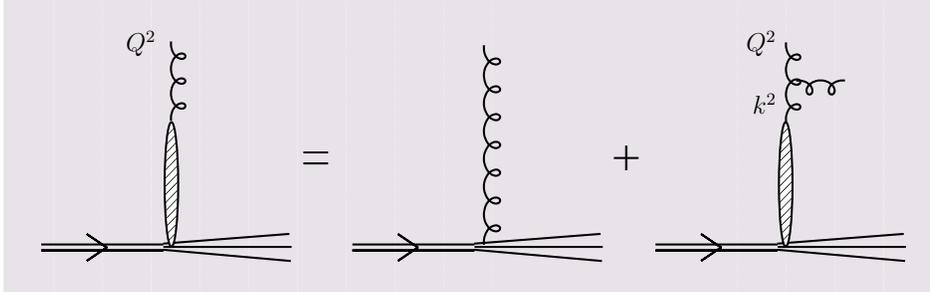}}
    \caption{Graphical depiction of the integral equation \eqref{eq:inteqn}}
    \label{fig:inteqn}
  \end{center}
\end{figure}

\begin{figure}
  \begin{center}
    \epsfig{file=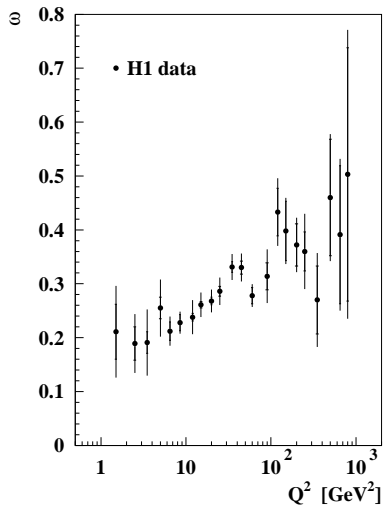,width=0.4\textwidth}
  \end{center}
  \caption{The effective power of the rise of $F_2$ \cite{H1omeff};
    $x$ and $Q^2$ values are correlated so that higher $Q^2$ also
    means higher $x$.}
  \label{fig:h1omeff}
\end{figure}

Our \textsc{DL} series happens to be related to the series for the
modified $I_0$ Bessel function \cite{AbSt}. Using the asymptotic
expansion for $I_0$ gives us the result that
\begin{equation}
  Q^2 \cF(x,Q^2) \sim \exp \left[ 2 \sqrt{\asb \ln \frac1x \ln
      \frac{Q^2}{Q_0^2}}\right] \,.
\end{equation}
However it is useful to develop a more general method of summation,
one which will be applicable also later on. Accordingly, we formulate
the problem as an integral equation
\begin{equation}
  \label{eq:inteqn}
  \cF(x,Q^2) = \cF^{(0)}(x,Q^2) + \asb \int^1_x \frac{\d z}{z}
  \int^{Q^2} \frac{\d k^2}{Q^2} \cF\left(\frac{x}{z}, k^2\right),
\end{equation}
which is depicted graphically in figure~\ref{fig:inteqn}. It can be
diagonalised by taking Mellin transforms with respect to both $x$ and
$Q^2$,
\begin{equation}
  \label{eq:mln}
  \cF(x,Q^2) = \int \frac{\d \om}{2\pi i} x^{-\om}
               \int \frac{\d \ga}{2\pi i}
               \frac1{Q^2} \left(\frac{Q^2}{Q_0^2}\right)^{\ga}
               \cF_{\ga,\om}\,, 
\end{equation}
with the contours running parallel to the imaginary axis, giving
\begin{equation}\label{eq:DGLAPmln}
  \cF_{\ga,\om} = \cF_{\ga,\om}^{(0)}
  + \asb \int_0^1 \frac{dz}{z} z^{\om} \int^{Q^2} \frac{dk^2}{Q^2}\,
  \frac{Q^2}{k^2} 
  \l( \frac{k^2}{Q^2} \r)^\ga \cF_{\ga,\om}
  \,=\, \cF_{\ga,\om}^{(0)} + \frac{\asb}{\om \ga}
  \cF_{\ga,\om}\,.
\end{equation}
The pole in $\ga$ is conjugate to the \textsc{DGLAP} logarithm of
$Q^2$ and the pole in $\om$ conjugate to the logarithm of $x$.
Eq.~\eqref{eq:DGLAPmln} is easily solved:
\begin{equation}
  \label{eq:DLmeln}
 \cF_{\ga,\om} = \frac{\om \cF_{\ga,\om}^{(0)}}{\om -
   \frac{\asb}{\ga}}\,.
\end{equation}
With the initial condition \eqref{eq:Finit}, we have
$\cF_{\ga,\om}^{(0)}=1/\om\ga$.  The inverse Mellin transform with
respect to $\om$ is carried out by closing the $\om$ contour to the
left in eq.~\eqref{eq:mln}, leaving us with
\begin{equation}
  \label{eq:DL1mln}
  Q^2 \cF(x,Q^2) =  \int \frac{\d \ga}{2\pi i}
\, x^{-\frac{\asb}{\ga}} 
\left(\frac{Q^2}{Q_0^2}\right)^{\ga}
\cdot \frac{1}{\ga}\,.
\end{equation}
The integrand has a saddle-point at
\begin{equation}
  \label{eq:DLsdl}
  \gabar = \sqrt{\frac{\asb \ln 1/x}{\ln Q^2/Q_0^2}}
\end{equation}
and a saddle-point evaluation of the integral gives
\begin{equation}
  \label{eq:DLres}
  Q^2 \cF(x,Q^2) \simeq \frac{1}{2}
  \left(\frac{1}{\pi^2 \asb \ln \frac1x \ln \frac{Q^2}{Q_0^2}}\right)^{1/4}
\exp \left[ 2 \sqrt{\asb \ln \frac1x \ln
      \frac{Q^2}{Q_0^2}}\,\right].
\end{equation}
This result was first obtained twenty-five years ago by De Rujula
\etal \cite{DLorig}. Its main feature is that the gluon distribution
rises at small $x$, with an effective power
\begin{equation}
  \label{eq:DLomeff}
  \om_\eff \simeq \sqrt{\frac{\asb \ln 1/x}{\ln Q^2/Q_0^2}}\,,
\end{equation}
which decreases as one moves towards smaller $x$ values, and increases
towards higher $Q^2$. The H1 data for $\om_\eff$ shown in
figure~\ref{fig:h1omeff} illustrate precisely this trend.
Detailed comparisons of \textsc{DGLAP}-induced rises of $F_2$, including full
splitting functions and a running coupling, such as those performed by
Gl\"uck, Reya and Vogt \cite{GRV} and Ball and Forte \cite{BF}, show
remarkably good agreement with nearly all the small-$x$ $F_2$
structure function data, even down to $Q^2\simeq 4$~GeV$^2$.

\subsection{BFKL}

The above arguments are relevant in a limit where both $1/x$ and
$Q^2/Q_0^2$ are large, \ie when we have strong ordering in both
longitudinal and transverse momenta. But when the ends of the chain
have similar transverse momenta, $Q^2 \simeq Q_0^2$, there is no
longer any reason for transverse momenta along the chain to be
ordered. Double logs no longer dominate the cross section and we have
to sum all leading (single) logarithms (\textsc{LL}) of $x$,
\begin{equation}
\label{eq:singlog}
  \l(\asb \ln \frac{1}{x}\r)^n.
\end{equation}
This is done by the \textsc{BFKL} equation \cite{BFKL}, which can be
derived in a number of ways.  Since these notes are intended only as
an introduction to the \textsc{BFKL} equation, rather than engaging in
its derivation we will try to deduce its main characteristics from
simple physical arguments.\footnote{For the interested reader one of
  the simplest full derivations is perhaps to be found within the
  dipole formulation \cite{dipoleBFKL}. A wide ranging introduction
  and discussion of many aspects of BFKL physics can be found in
  \cite{FRbook}.}

In the previous section we had the following integral equation for the
gluon density \eqref{eq:inteqn}:
\begin{equation}
  \label{eq:inteqnrepeat}
  \cF(x,Q^2) = \cF^{(0)}(x,Q^2) + \int \frac{\d z}{z}
  \int \d k^2 K(Q^2,k^2)\, \cF\left(\frac{x}{z}, k^2\right),
\end{equation}
where the \textsc{DGLAP} kernel $K$ was just
\begin{equation}\label{eq:DGLAPkern}
  K(Q^2,k^2) = \frac{\asb}{Q^2} \Theta(Q^2-k^2) \,,\qquad\qquad 
  \mbox{valid for }Q^2 \gg k^2\,.
\end{equation}
Since the \textsc{BFKL} kernel should be valid for any ratio of
transverse scales it must have the same limit for $Q^2\gg k^2$, and
additionally correctly treat situations in which $Q^2$ is of the same
order as, or much smaller than $k^2$. We can deduce its form in the
limit $k^2\gg Q^2$ by the following argument.  The scattering of a big
object off a small one, or of a small object off a big one, must have
the same cross section, and both situations must be correctly
described by the BFKL resummation. Therefore the \textsc{BFKL} kernel
must be symmetric under the interchange of $Q^2$ and $k^2$.  So when
$k^2 \gg Q^2$ (the anti-\textsc{DGLAP}, or anti-collinear limit) we
have
\begin{equation}
  K(Q^2,k^2) = \frac{\asb}{k^2}  \,,\qquad\qquad 
  \mbox{valid for }k^2 \gg Q^2\,.
\end{equation}
If we approximate the full kernel just by its collinear and
anti-collinear limits, then we have
\begin{equation}\label{eq:Kcoll}
  K^{\coll}(Q^2,k^2) = \asb\cdot \l(\frac{\Theta(Q^2-k^2)}{Q^2} +
  \frac{\Theta(k^2-Q^2)}{k^2}\r).
\end{equation}
Following the treatment of the previous section, we will need its
Mellin transform,
\begin{equation}\label{eq:chicoll}
  \chi^{\coll}(\ga) = \frac{1}{\ga} + \frac{1}{1-\ga}\,,
\end{equation}
where, by convention, the leading factor of $\asb$ has been left out;
$\chi(\ga)$ is usually referred to as the \emph{characteristic
  function} of the system. The $1/\gamma$ term was present also in the
pure DGLAP case, and comes from the collinear limit. The $1/(1-\ga)$
term comes from the anti-collinear limit.  The symmetry
$\ga\leftrightarrow 1-\ga$ is a direct consequence of the symmetry
under the exchange of the two transverse scales. This can be seen
explicitly from the definition of the Mellin transform,
eq.~\eqref{eq:mln}, where
\begin{equation}
  \frac1{Q^2} \left(\frac{Q^2}{Q_0^2}\right)^{\ga}
  =   
  \frac1{Q_0^2} \left(\frac{Q_0^2}{Q^2}\right)^{1-\ga}.
\end{equation}

\begin{figure}[htbp]
  \begin{center}
    \epsfig{file=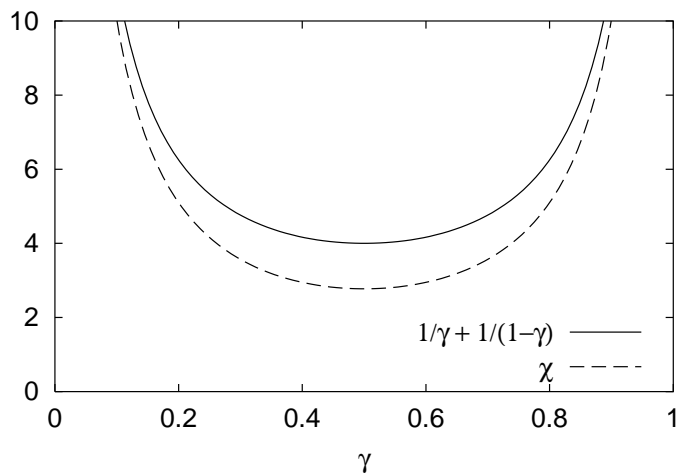,width=0.7\textwidth}
    \caption{The full and collinearly-approximated characteristic functions.}
    \label{fig:chillrg}
  \end{center}
\end{figure}

What we have neglected in our collinear $+$ anti-collinear
approximation is the correct treatment of the kernel for $k$ of the
same order as $Q$. This is given by the full \textsc{BFKL} kernel
\cite{BFKL} (with integration measure $d^2k/\pi$ because the azimuthal
integration now matters):
\begin{equation}
  \label{eq:BFKLkern}
  K(Q^2,k^2) = \asb \cdot \left(\frac{1}{|\vec Q - \vec k|^2} -
    \delta(Q^2 - k^2) 
  \int^k \frac{\d^2 q}{\pi q^2}\right)\,.
\end{equation}
Its Mellin transform (again leaving out the overall factor of $\asb$)
is
\begin{equation}
  \label{eq:BFKLchi}
  \chi(\ga) = 2\psi(1) - \psi(\ga) - \psi(1-\ga),
\end{equation}
where $\psi(x) = \d \ln \Gamma(x)/\d x$. Noting that $-\psi(\ga) =
1/\ga + \cO{1}$ for small $x$, we see that the full $\chi(\ga)$ has
the same polar structure around $\ga=0$ and $\ga=1$ as our
approximation \eqref{eq:chicoll} reflecting the fact that the
collinear and anti-collinear limits are the same. The two
characteristic functions are shown in fig.~\ref{fig:chillrg}, which
illustrates their very similar shapes: they differ by little more than
a constant.

\begin{figure}[htbp]
  \begin{center}
    \epsfig{file=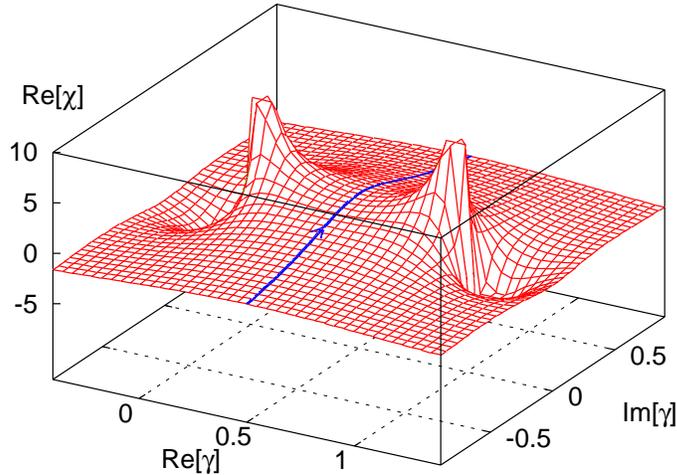,width=0.7\textwidth}
    \caption{The real part of the \textsc{LL} \textsc{BFKL}
      characteristic function in the complex plane and the integration
      contour for the inverse Mellin transform.}
    \label{fig:chill}
  \end{center}
\end{figure}

The procedure for obtaining the BFKL cross section is analogous to
that used in the \textsc{DGLAP} case, with $1/\ga$ replaced by
$\chi(\ga)$.  We start with the Mellin-transformed integral equation
\begin{equation}\label{eq:BFKLmln}
  \cF_{\ga,\om} = \cF_{\ga,\om}^{(0)} + \frac{\asb}{\om} \chi(\ga)
  \cF_{\ga,\om}\,,
\end{equation}
and solve for $\cF_{\ga,\om}$:
\begin{equation}
  \label{eq:cFBFKLmell}
  \cF_{\ga,\om} = \frac{\om \cF_{\ga,\om}}{\om - \asb\chi(\ga)}
\end{equation}
The inverse Mellin transform with respect to $\om$ is once again
trivial and gives
\begin{equation}
  \label{eq:LL1mln}
  \cF(x,Q^2) = \int \frac{\d \ga}{2\pi i} \exp\left[\asb\chi(\ga)
    \ln \frac1x  + 
  \ga \ln \frac{Q^2}{Q_0^2}\right] \cdot \ombar \cF^{(0)}_{\ga,\ombar}
\qquad \ombar = \asb\chi(\ga)\,.
\end{equation}
The real part of the characteristic function $\chi(\ga)$, together
with the integration contour are shown in the complex plane in
figure~\ref{fig:chill}. For large $\ln 1/x$, the integrand in
\eqref{eq:LL1mln} is dominated by behaviour of $\chi$ and has a
saddle-point close to $\ga=1/2$, which causes the gluon distribution
to grow as
\begin{equation}
  \label{eq:BFKLsdlres}
  \cF(x,Q^2) \simeq \frac{x^{-\asb \chi\l(\frac12 \r)}}{\sqrt{2\pi \,\asb
        \,\chi''\!\l(\frac12\r) \ln \frac1x }}  \cdot \frac{1}{Q Q_0}\,,
\end{equation}
where $\chi''$ is the second derivative of $\chi$ with respect to
$\ga$.  Thus the gluon distribution (and in general, high-energy cross
sections) should grow as a power of $x$ determined by the minimum
value of $\chi(\ga)$, which is $\chi(1/2)=4\ln2$. For the
phenomenologically reasonable value of $\asb\simeq\as=0.2$ this gives
a power of about $0.5$.

It suffices to look back at figure~\ref{fig:h1omeff} to see that this
is incompatible with the rise seen in the bulk of the structure
function data.  Some care is needed in interpreting this disagreement:
in considering the structure function data, we are trying to apply
perturbative \textsc{QCD} to a problem which is inherently
non-perturbative (the scale $Q_0^2$ does not satisfy our requirement
$Q_0^2\gg\Lambda^2$). However \textsc{BFKL} also predicts scaling
violations of the $F_2$ structure function \cite{BFKLScaling}, and
this prediction can be shown not to depend on the properties of the
non-perturbative region \cite{CCS,CCS2}.  Essentially, regardless of
the input distribution, the scaling violations quickly lead to a
structure function which rises with a power $ 4\ln 2 \asb$ and so is
incompatible with the data \cite{BFBFKL}.

\begin{figure}[tbp]
  \begin{center}
    \epsfig{file=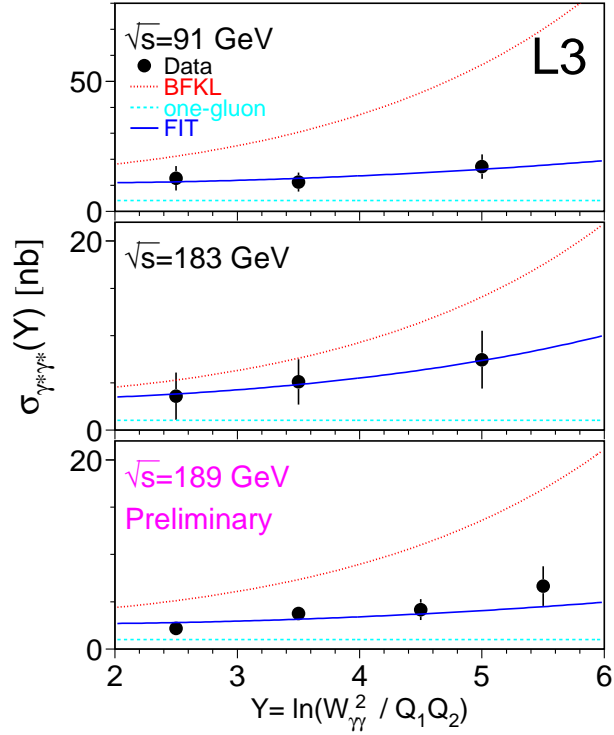,width=0.65\textwidth}
    \caption{The cross section for $\gaga$ collisions as
      measured by the L3 collaboration \cite{L3,Maneesh}. The mean
      $Q^2$ values for the three energies are $3.5$, $14$ and
      $14.5\gev^2$ respectively.}
    \label{fig:l3}
  \end{center}
\end{figure}

There exist other, theoretically cleaner, tests of \textsc{BFKL}.
Generally they involve selecting a process with two hard hadronic
probes, such as jets or a virtual photon, separated by a large
rapidity (or equivalently having a large centre-of-mass energy). The
requirement that both probes be hard ensures that one can reasonably
apply perturbation theory\footnote{Though not strictly the subject of
  this presentation, an exposition of BFKL physics would be incomplete
  without at least some mention of \emph{diffusion}. Because
  transverse momenta are not ordered, small-$x$ evolution leads to a
  random walk in $\ln k_t$. The mean width of this random walk ---
  diffusion --- increases as $\sqrt{\ln s}$, and at very large $s$
  eventually enters into the non-perturbative region. Thus, no matter
  how large the transverse scales of the scattering objects, there is
  always an energy beyond which perturbation theory loses its
  predictive power.} (unfortunately it generally also makes the
experimental measurement much harder). A nice example of such a test
is the collision of two virtual photons as measured recently by the L3
\cite{L3} and \textsc{OPAL} \cite{OPAL} collaborations. The L3 data
are shown in figure~\ref{fig:l3}. The data are significantly higher
than the one-gluon estimate (\ie the prediction without \textsc{BFKL}
resummation). On the other hand the \textsc{LL} \textsc{BFKL}
predictions clearly overshoot the data.  The L3 collaboration perform
a fit to the data in order to determine the power of the high-energy
growth, and quote a preliminary result of $0.29\pm0.025$ for scales in
the range of $3.5$ to $14.5\gev^2$ \cite{Maneesh}.\footnote{This
  result should probably be interpreted with some caution because the
  formula used to carry out the fit assumes the \textsc{LL}
  normalisation with four light-quark flavours (whereas both the
  \textsc{NLL} corrections and the charm mass probably have a
  significant effect on the normalisation).  A fit leaving the
  normalisation as a free parameter leads to a similar power but with
  a much larger error. One should bear in mind that because of the
  `limited' energies at \textsc{LEP}, the $Q^2$ values (between $3.5$
  and $14.5$~GeV$^2$) are probably on the border of the region that
  can be considered perturbative.}

The same conclusion of incompatibility with \textsc{LL} \textsc{BFKL}
comes out from looking at the interactions between a jet and a virtual
photon in \textsc{DIS} \cite{FJ}, a measurement referred to as the
forward-jet cross section, because of the position of the jet in the
detector.

\section{Next-to-leading corrections}
\label{sec:nll}

All along, while the various experimental tests of \textsc{LL}
\textsc{BFKL} were being carried out and refined, the calculation of
the next-to-leading logarithmic corrections to \textsc{BFKL} was in
progress.  The next-to-leading terms are those suppressed by a power
of $\as$ relative to the \textsc{LL} series:
\begin{equation}
  \as (\as \ln s)^n\,.
\end{equation}
In terms of the notation developed so far, this corresponds to working 
out the \textsc{NLL} corrections to the characteristic function $\chi$, \ie
finding $\chi_1$, where
\begin{equation}
  \label{eq:chigen}
  \asb \chi(\ga) = \asb\chi_0(\ga) + \asb^2 \chi_1(\ga) + \cO{\asb^3}\,.
\end{equation}
The determination of $\chi_1$ took close to ten years \cite{NLL}, and
was completed quite recently \cite{FL,CC98}.

Rather than trying to reproduce parts of that derivation, we will
adopt the same approach that was used in the previous section, namely
to deduce the structure of the \textsc{NLL} corrections through a
study of the collinear limit and symmetry requirements. 
This will translate to determining the divergences around $\ga=0$ and
$\ga=1$.

We will examine three main contributions: those from the running
coupling, the non-singular (at small $z$) part of the splitting
functions and the choice of energy scale.

\subsection{Running coupling}
\label{sec:nll-run}

The \textsc{QCD} coupling runs as
\begin{equation}
  \asb(k^2) = \frac{\asb(Q^2)}{1 + b\, \asb(Q^2) \ln \frac{k^2}{Q^2}}\,,
\end{equation}
where $b = 11/12 - \nf/6$. What sort of higher order contribution will
this lead to? The \textsc{DGLAP} equations tell us that in the
right-hand graph of fig.~\ref{fig:inteqn}, when $Q^2 \gg k^2$ the
correct scale for the coupling is $Q^2$. By symmetry, when $k^2 \gg
Q^2$, the correct scale is $k^2$ --- \ie in the collinear limit the
correct scale is the larger of the two scales involved. So our
collinear approximation for the kernel, eq.~\eqref{eq:Kcoll}, becomes
\begin{equation}
  \label{eq:Kcollrun}
    K^{\coll}(Q^2,k^2) = \asb(Q^2)\frac{\Theta(Q^2-k^2)}{Q^2} +
  \asb(k^2)\frac{\Theta(k^2-Q^2)}{k^2}\,.
\end{equation}
The Mellin transform of the first term just gives $\asb(Q^2)/\ga$, as
before. For the second term, we re-express $\asb(k^2)$ in terms of
$\asb(Q^2)$ in order to extract a factor of $\asb(Q^2)$ in front of
the whole result. Expanding to second order, and taking the Mellin
transform, gives
\begin{equation}
  \int_{Q^2}  \frac{\d k^2}{k^2} \l(\asb(Q^2) -
  b\asb^2\ln\frac{k^2}{Q^2}\r)  
  \,  \frac{Q^2}{k^2}\l(\frac{k^2}{Q^2}\r)^\ga
  = \frac{\asb(Q^2)}{1-\ga} - \frac{b \asb^2}{(1-\ga)^2}\,,
\end{equation}
which is just the anti-collinear part of our \textsc{LL} result plus a
running-coupling \textsc{NLL} contribution
\begin{equation}
  \label{eq:chi1b}
  \chi_1^b = -\frac{b}{(1-\ga)^2}\,.
\end{equation}
The lack of symmetry $\ga \leftrightarrow 1-\ga$ is due to our choice
to extract an asymmetric factor of $\asb(Q^2)$ in front of the answer.

What is the uncertainty on our collinear approximation for $\chi_1^b$?
The scheme of $\asb$ is not defined, corresponding to an uncertainty
proportional to $\chi_0$. Nor do we a priori know the correct scale
for branchings when $k^2$ and $Q^2$ are of the same order. So the
overall uncertainty is a function with at most single poles at $\ga=0$
and $\ga=1$.

\subsection{Splitting function}
\label{sec:nll-split}

\begin{floatingfigure}{0.3\textwidth}
  \hspace{-1.7cm}\begin{picture}(0,0)%
\epsfig{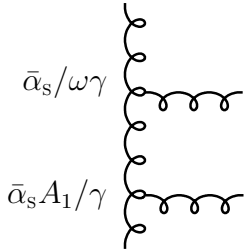}%
\end{picture}%
\setlength{\unitlength}{4144sp}%
\begingroup\makeatletter\ifx\SetFigFont\undefined%
\gdef\SetFigFont#1#2#3#4#5{%
  \reset@font\fontsize{#1}{#2pt}%
  \fontfamily{#3}\fontseries{#4}\fontshape{#5}%
  \selectfont}%
\fi\endgroup%
\begin{picture}(1979,1516)(-162,-1413)
\put(991,-1113){\makebox(0,0)[rb]{\smash{\SetFigFont{12}{14.4}{\familydefault}{\mddefault}{\updefault}$\asb A_1/\ga$}}}
\put(991,-437){\makebox(0,0)[rb]{\smash{\SetFigFont{12}{14.4}{\familydefault}{\mddefault}{\updefault}$\asb/\om\ga$}}}
\end{picture}

  \caption{A sequence of a small-$x$ and a non-small-$x$
    branching.\vspace{0.5cm}}
  \label{fig:twoemsn}
\end{floatingfigure}
In section~\ref{sec:LL} we used only the part of the gluon splitting
function that is singular at small $z$. At \textsc{NLL}, we need to
include the full splitting function \eqref{eq:pgg}. Its Mellin
transform (with respect to $x$) is
\begin{equation}
  \label{eq:Pgg}
  P_{gg}^{\om} = \frac1\om + A_1(\om)\,,
\end{equation}
where (for $\nf=0$)
\begin{equation}
  \label{eq:A1}
  A_1(\om) 
  = -\frac{11}{12} + \cO{\om}\,.
\end{equation}
\noindent To get the \textsc{NLL} correction we consider a sequence of two
collinear branchings, fig.~\ref{fig:twoemsn}, where one of the
branchings is a small-$x$ branching, giving a factor $\asb/\om\ga$ and
the other is a non-small-$x$ branching, giving a factor $\asb
A_1/\ga$. Remembering that convolutions in $x$ and $k^2$ space
translate to products in the $\om,\gamma$ Mellin transform space, our
integral equation \eqref{eq:BFKLmln} receives a contribution
\begin{equation}
  \frac{\asb}{\om} \frac{\asb A_1}{\ga^2} \cF_{\ga,\om}\,.
\end{equation}

\noindent There is a corresponding term for a pair of anti-collinear
branchings, so that the splitting-function contribution to $\chi_1$ is
\begin{equation}
  \label{eq:chi1A1}
  \chi_1^{A_1}(\ga) = \frac{A_1}{\ga^2} + \frac{A_1}{(1-\ga)^2}\,,
\end{equation}
where $A_1=A_1(0) = -11/12$. Actually this is only the
$\nf$-independent part. For non-zero $\nf$ there are contributions
coming from the $\nf$-dependence of $P_{gg}$ and from diagrams
involving the convolution of $P_{gq}$ and $P_{qg}$.

As in the running coupling case we have an uncertainty on this result,
which can arise for example from the combination of a collinear and an
anti-collinear branching, and thus is once again at the level of a
function with at most single poles at $\ga=0$ and $\ga=1$.

\vspace{0.6cm}

\subsection{Energy scale terms}
\label{sec:s0}

A more subtle source of \textsc{NLL} corrections comes from the so-called
energy-scale terms. At leading order one resums terms
\begin{equation}
 \l( \asb \ln \frac{s}{s_0}\r)^n\,,
\end{equation}
where $s_0$ can be chosen arbitrarily. Changing $s_0$ is equivalent to
introducing a whole set of higher order terms.  For a symmetric
treatment (with respect to $Q$ and $Q_0$), a natural choice is
$s_0=Q_0 Q$. Let us then consider what happens when $Q\gg Q_0$.  As we
obtained in section~\ref{sec:dis}, the leading terms are
\begin{equation}
  \frac{1}{(n!)^2}\l( \asb \ln \frac{s}{Q Q_0} \ln \frac{Q^2}{Q_0^2}
  \r)^n\,.
\end{equation}
But for \textsc{DIS}-like situations, $Q\gg Q_0$, we usually express
our results as a function of $x = Q^2/s$ and $Q^2/Q_0^2$. Let us do
that for the $n=2$ term:
\begin{equation}
  \label{eq:dlord2}
  \frac{1}{4}\l( \asb \ln \frac{s}{Q Q_0} \ln \frac{Q^2}{Q_0^2}
  \r)^2 = 
  \frac{1}{4}\l( \asb \ln \frac{1}{x} \ln \frac{Q^2}{Q_0^2}
  \r)^2 + \frac{1}{4} \asb^2 \ln \frac{1}{x} \ln^3 \frac{Q^2}{Q_0^2}
  + \textsc{NNLL}\,.
\end{equation}
Of particular interest is the second term on the \textsc{RHS} because
it has more collinear logs than powers of $\asb$ (it contains a double
collinear logarithm for a single power of $\asb$). But we know from
renormalisation group constraints that the cross section written as a
function of $x$ and $Q^2/Q_0^2$ contains at most as many collinear
logs as powers of $\asb$. Therefore for the result to be consistent
with the renormalisation group, the next-to-leading corrections must
be such as to cancel the second term on the \textsc{RHS} of
\eqref{eq:dlord2}, \ie they must contain a term
\begin{equation}
  \label{eq:lncubed}
  -\frac{1}{4} \asb^2 \ln s \ln^3 \frac{Q^2}{Q_0^2}\,.
\end{equation}
In Mellin transform space this will correspond to a contribution to
$\chi_1$ which is proportional to $1/\ga^3$. To obtain its coefficient
it is not sufficient just to take the Mellin transform of
\eqref{eq:lncubed}, because not all of the correction exponentiates
(\ie should be incorporated into $\chi$) --- for example some of it is
to be associated with the initial condition.

Instead, now that we know what kind of answer to expect, let us
directly consider the problem in Mellin-transform space.  We start
with a result written for energy scale $Q Q_0$ (\cnf eqs.\ 
\eqref{eq:mln} and \eqref{eq:cFBFKLmell}),
\begin{equation}
  \label{eq:esclqqz}
  \cF(s,Q^2) = \int \frac{\d \om}{2\pi i}
               \int \frac{\d \ga}{2\pi i}
               \l(\frac{s}{Q Q_0}\r)^\om
               \frac1{Q^2} \left(\frac{Q^2}{Q_0^2}\right)^{\ga}
               \frac{\om \cF^{(0)}_{\ga,\om} }{\om - \asb \chi(\ga)}\,,
\end{equation}
and then note that since
\begin{equation}
\l(\frac{s}{Q Q_0}\r)^\om \left(\frac{Q^2}{Q_0^2}\right)^{\ga} = 
\l(\frac{s}{Q^2}\r)^\om \left(\frac{Q^2}{Q_0^2}\right)^{\ga+\hom},
\end{equation}
rewriting \eqref{eq:esclqqz} with energy scale $s_0=Q^2$ is just
equivalent to shifting $\ga \to \ga-\hom$ in $\chi$ and $\cF^{(0)}$:
\begin{equation}
  \label{eq:esclqsqr}
  \cF(s,Q^2) = \int \frac{\d \om}{2\pi i}
               \int \frac{\d \ga}{2\pi i}
               \l(\frac{s}{Q^2}\r)^\om
               \frac1{Q^2} \left(\frac{Q^2}{Q_0^2}\right)^{\ga}
               \frac{\om \cF^{(0)}_{\ga-\hom,\om}}{\om - \asb
                 \chi(\ga-\hom)}\,. 
\end{equation}
If we now expand $\chi(\ga-\hom)$ in powers of $\asb$, recursively
using the relation $\om=\asb\chi$ we get
\begin{equation}
  \chi(\ga-\hom) = \chi(\ga) - \frac{\asb\chi   \chi'}{2} +
  \cO{\asb^2}\,.
\end{equation}
In the collinear limit ($\ga\to 0$), since $\chi(\ga)$ goes as
$1/\ga$, the $\cO{\asb}$ piece has the behaviour
\begin{equation}
  -\frac{\asb\chi   \chi'}{2} \simeq \frac{\asb}{2\ga^3}\,.
\end{equation}
This is the analogue of the $\ln^3 Q^2$ term seen earlier
\eqref{eq:dlord2} and it must be subtracted from $\chi$ at scale
$s_0=Q Q_0$ in order for the collinear limit with energy scale $Q^2$
to be free of unwanted double collinear logs. There is an analogous
$1/2(1-\ga)^3$ piece to be subtracted for the anti-collinear limit to
be correct (\ie free of double anti-collinear logs for energy scale
$s_0=Q_0^2$).  Overall therefore we have the following \textsc{NLL}
energy-scale corrections (for $s_0=Q Q_0$):
\begin{equation}
  \label{eq:chi1s0}
  \chi_1^{s_0} = -\frac{1}{2\ga^3} - \frac{1}{2(1-\ga)^3}\,.
\end{equation}
As was the case for the running coupling and splitting function terms,
our analysis leaves us with an uncertainty which amounts to a function
with at most single poles at $\ga=0$ and $\ga=1$.

\subsection{Putting things together}
\label{sec:together}

Putting together eqs.~\eqref{eq:chi1b}, \eqref{eq:chi1A1} and
\eqref{eq:chi1s0} gives us the following answer for the
collinearly-enhanced part of the \textsc{NLL} corrections ($\nf=0$):
\begin{equation}
  \label{eq:chi1coll}
  \chi_1^{\coll}(\ga) = \frac{A_1}{\ga^2} 
  + \frac{A_1-b}{(1-\ga)^2} -\frac{1}{2\ga^3} - \frac{1}{2(1-\ga)^3}\,,
\end{equation}
with $A_1=-11/12$. Our collinear approximation guarantees the
correctness of the coefficients of the cubic and quadratic divergences
at $\ga=0$ and $\ga=1$.

The true \textsc{NLL} corrections as assembled in \cite{FL,CC98} are,
in the $\MSbar$ scheme and for $\nf=0$ (the $\nf$ dependence turns out
to be small)
\begin{multline}
  \label{eq:chi1true}
  \chi_1(\ga) = 
{ - \frac{\pi ^2\cos(\pi \ga )}{4\sin^2(\pi \ga
)(1-2\ga )}\left( 3+ \frac{2+3\ga
(1-\ga )}{(3-2\ga )(1+2\ga )}\right) }
\\
{-\frac{b}{2}\left(\chi_0^2(\ga ) - \psi'(\ga) +
    \psi'(1-\ga)\right)} 
{ +\frac{\psi ^{\prime \prime }(\ga )}{4} +
  \frac{\psi ^{\prime \prime }(1-\ga )}{4}}
\\{
+ \left(\frac{67}{36} - \frac{\pi^2}{12}\right) \chi_0 (\ga )
 +\frac{3}{2}\zeta (3)
+ \frac{\pi ^3}{4\sin (\pi \ga )}-\phi (\ga )}\,,
\end{multline}
where
\begin{equation}
  \label{eq:chi1phi}
  \phi (\gamma )=\sum_{n=0}^\infty (-1)^n\left[ \frac{\psi (n+1+\gamma
    )-\psi (1)}{(n+\gamma )^2}+\frac{\psi (n+2-\gamma )-\psi
    (1)}{(n+1-\gamma )^2}\right]\,.
\end{equation}
It is possible to make a direct identification between parts of
\eqref{eq:chi1true} and \eqref{eq:chi1coll} in terms of the
coefficients of the double and triple poles. The first line of
\eqref{eq:chi1true} is identifiable with the $A_1$ piece of
\eqref{eq:chi1coll} and so originates from the splitting function. The
running-coupling dependence enters through the first term on the
second line of \eqref{eq:chi1true}, while the energy-scale dependent
piece is formed by the last two terms of that line. The remaining
terms are free of double and triple poles. Of these terms, so far only
the first one on the third line of \eqref{eq:chi1true} has been
understood: it is associated with the fact that the natural scheme for
processes involving soft gluons is the \textsc{CMW} or
gluon-bremsstrahlung scheme \cite{CMW}. When writing an answer in the
$\MSbar$ scheme this leads to a correction term which is
$(67/36-\pi^2/12)\asb$ times the leading order result.

\begin{figure}[htbp]
  \begin{center}
    \epsfig{file=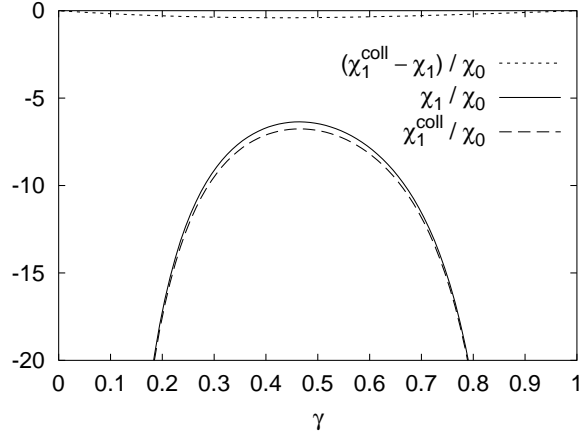,width=0.6\textwidth}
    \caption{A comparison of our collinear approximation and the full
      result for $\chi_1$; $\nf=0$.}
    \label{fig:chi1coll}
  \end{center}
\end{figure}

Figure~\ref{fig:chi1coll} shows the full $\chi_1$ together with our
collinear approximation. There is a remarkable similarity between
them: in the range $0<\ga<1$ they never differ by more than $7\%$.
Possible reasons for the surprisingly good agreement will be discussed
later, in section \ref{sec:resum}.

\subsection{Consequences of the \textsc{NLL} corrections}
\label{sec:nloprobs}

\begin{figure}[htbp]
  \begin{center}
    \epsfig{file=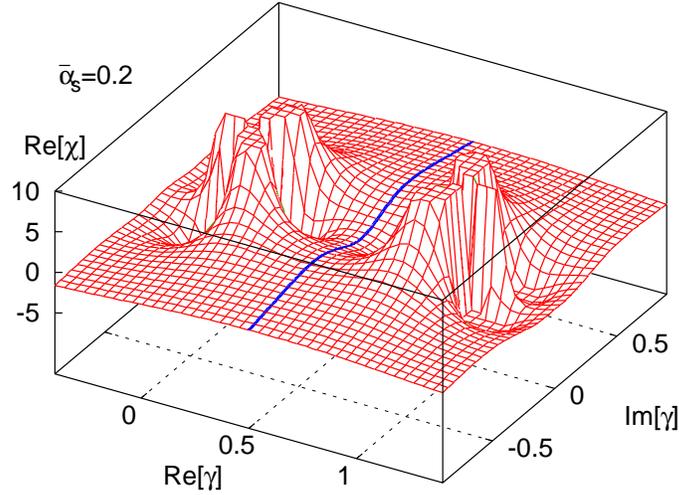,width=0.7\textwidth}
    \caption{The real part of the \textsc{LL} $+$ \textsc{NLL}
      characteristic function 
      for $\asb=0.2$, together with the integration contour, in the
      complex plane.} 
    \label{fig:chinll}
  \end{center}
\end{figure}

Figure~\ref{fig:chi1coll} shows that the \textsc{NLL} corrections to $\chi$ are
very large. For $\ga=1/2$ we have (again for $\nf=0$)
\begin{equation}
  \chi(1/2) = \chi_0(1/2)(1 - 6.47 \asb)\,.
\end{equation}
Thus for $\asb=0.2$, the predicted power is negative, at about
$-0.16$, which is no more in agreement with the data than the leading
power. 

What is even more worrisome is that the structure of the
characteristic function changes radically. In figure~\ref{fig:chill},
there is a single saddle-point at $\ga=1/2$, \ie on the real axis.
With the \textsc{NLL} contributions included, there are now two
saddle-points, at complex values of $\ga$ which we will call
$\gabar,\gabar^*$ \cite{Ross}.  Since the cross section goes as
\begin{equation}
  \sigma(s,Q^2,Q^2_0) \sim \l(\frac{s}{Q Q_0}\r)^{\asb\chi(\gabar)} 
  \frac{1}{Q^2}\l(\frac{Q^2}{Q_0^2}\r)^\gabar + \gabar \leftrightarrow 
  \gabar^*,
\end{equation}
the fact that $\gabar$ is complex means that the cross section
oscillates as a function of $\ln Q^2/Q^2_0$ (it remains real because
of the contribution from the complex-conjugate saddle point).  This
behaviour occurs as long as $\chi$ has a negative second derivative at
$\ga=1/2$, which is the case for $\asb \gtrsim 0.05$. In other words
there exists no phenomenologically accessible domain in which the
inclusion of the \textsc{NLL} corrections gives a sensible result.

\section{Beyond NLL}
\label{sec:resum}

The solution is bound to lie with higher orders. Shortly after
preliminary results on $\chi_1$ had appeared, it was suggested that
stable predictions might be obtained by inclusion of the \textsc{NNLL}
and \textsc{NNNLL} terms \cite{blumvogt}. But remembering that the
\textsc{LL} calculation took about a year, and the \textsc{NLL}
calculation ten years, a reasonable estimate for the time to calculate
the \textsc{NNLL} terms might lie somewhere between an arithmetic (19
years) and a geometric (100 years) extrapolation. Even were these
contributions to be calculated, there is actually no guarantee that
the resulting series would converge for the values of $\asb$ of
interest!

So the only option left is to try and \emph{guess} the higher-order
terms and then to resum them (we are now talking about the resummation
of a resummation). The question is whether there is some reliable way
of guessing them. Various approaches have been investigated
\cite{Schmidt,BFKLP,GPS,CCS,FRS}. I here will advocate a method
closely related to that used in the previous section to estimate the
\textsc{NLL} corrections --- namely a method based on the study of the
collinear limit \cite{GPS,CCS}.

We have already seen in the previous section that a study of the
collinear limit is a powerful tool. For the \textsc{NLL} characteristic
function it gave us the cubic and quadratic divergences at $\ga=0$ and
$\ga=1$, and in the range $0<\ga<1$ reproduced the full answer to
remarkably good accuracy. How come?

There is a temptation to argue that since $\ga=1/2$ is a moderately
small number, one can legitimately carry out an expansion in powers of
$\ga$ and $1-\ga$ (including the first two terms in the expansion for
$\chi_0$, \ie the two poles and a constant, also seems to do quite
well, reproducing the full answer to within about $8\%$).  A slightly
better motivated argument might be the following. For $\ga=1/2$, the
transverse momentum integrals in the Mellin transform converge quite
rapidly and so one might not expect a collinear approximation to work
too well. However at higher orders, pieces of the transverse momentum
integrals are accompanied by logarithms of transverse momentum. These
have the effect of shifting the dominant part of the integral out
towards more collinear regions, where the collinear approximation
itself becomes better.

So there are reasons to believe that collinearly-enhanced
contributions might give a significant part of the higher-order
corrections even beyond \textsc{NLL}.\footnote{This statement should
  really be restricted to those corrections that can be associated
  with a single ladder (referred to as $t$-channel iteration).
  Actually at \textsc{NNLL}, corrections arise associated with the
  presence of two ladders (the start of $s$-channel iteration), \ie
  saturation, or unitarity corrections. Our aim here is to understand
  the high-energy behaviour
  of a single ladder.} %
A more general justification for carrying out the collinear
resummation is that one wants to be able to use one's answer in the
collinear limit.  Since the collinear limit involves taking $\ga$
close to zero, where higher orders involve successively more divergent
terms, the only hope of a sensible answer there is a collinear
resummation.

The determination of the collinearly enhanced corrections can be
divided into two parts. The first deals with terms in the same class
as the $1/\ga^2$ terms in $\chi_1$, single collinear logarithms
originating from the splitting function and running coupling; the
second addresses the terms in the same class as the $1/\ga^3$ term in
the \textsc{NLL} result, namely double collinear logarithms. We will
consider only an outline of the method. The interested reader is
referred to the original references \cite{GPS,CCS} for the full
details.

\subsection{Single collinear logs --- running coupling and
  splitting function terms}
\label{sec:scl}

It is fairly straightforward to calculate the collinear N$^n$LL
contributions to the \textsc{BFKL} kernel from splitting function and
running coupling effects. One just takes diagrams such as
figure~\ref{fig:twoemsn} with an arbitrary number of non-small-$x$
emissions, inserting and expanding the appropriate running coupling
for each branching (the answer is given in \cite{CCS}). One is then
left with the tricky problem of resumming the resulting set of terms.

An equivalent approach essentially treats the small-$x$ and
non-small-$x$ branchings on a more similar footing \cite{CCS}. In
eq.~\eqref{eq:BFKLmln} we have a factor $\asb/\om$ coming from the
$1/z$ part of the $P_{gg}$ splitting function, and a factor of
$\chi(\ga)$ from the transverse structure of the branching. In the
collinear limit, we can replace $1/\om$ with the full splitting
function, $P_{gg}^\om$. So the collinear behaviour of the
$\asb\chi/\om$ factor becomes (for \textsc{DIS} energy scale,
$s_0=Q^2$)
\begin{equation}
  \label{eq:allcoll}
   \frac{\asb P_{gg}^\om }{\ga} = \frac{\asb(Q^2)}{\om} 
  \frac{1 + \om A_1(\om)}{\ga} \quad \Rightarrow \quad
   \chi \simeq \frac{1 + \om A_1(\om)}{\ga}
  \,,
\end{equation}
where we have used eq.~\eqref{eq:Pgg} for $P_{gg}^\om$. The correct
scale in the branching is $Q^2$ (the largest scale in the problem) so
there are no running coupling corrections.

The situation in the anti-collinear limit is very similar except for
an issue related to the running coupling: the appropriate scale for
the branching is not $Q^2$, but $k^2$ (referred to
fig.~\ref{fig:inteqn}). If we want to extract a factor of $\asb(Q^2)$
the difference of scales must be taken into account. It turns out
\cite{CCS} that this can be done at all orders by replacing $A_1$ with
$A_1-b$, so that the resummed anti-collinear structure is (with the
anti-DIS energy scale choice, $s_0=Q_0^2$)
\begin{equation}
  \label{eq:allantcoll}
   \frac{\asb(Q^2)}{\om} 
  \frac{1 + \om (A_1(\om) - b)}{1-\ga} \quad \Rightarrow \quad
   \chi \simeq \frac{1 + \om (A_1(\om) - b)}{1-\ga}\,.
\end{equation}
The reader can verify that substituting $\om=\asb\chi_0$ into the
expressions for $\chi$ in \eqref{eq:allcoll} and \eqref{eq:allantcoll}
reproduces the correct NLL collinear-enhanced terms.

\subsection{Double collinear logs --- energy-scale terms}
\label{sec:dcl}

We have just given resummed answers for the collinear and
anti-collinear behaviours of the kernel with \textsc{DIS} and
anti-\textsc{DIS} energy scale choices respectively. We really want
the answer for a common energy scale, say $s_0=Q Q_0$. We saw in
section~\ref{sec:s0}, that changes in $s_0$ introduce higher-order
double-collinear logs (the $1/\ga^3$ and $1/(1-\ga)^3$ terms). The
higher-order corrections had to be such that for energy scale $s_0 =
Q^2$ there were no such terms around $\ga=0$, and similarly around
$\ga=1$ for energy scale $s_0 = Q_0^2$. For $s_0=Q Q_0$, appropriate
double-collinear log counterterms had to be included in order to
satisfy the conditions for the other energy scales. One can work out,
order by order, the counterterms for higher kernels, but it soon gets
tedious. In any case one finds that the resulting series of terms is
divergent for reasonable values of $\as$.

The solution \cite{GPS} exploits the fact that a change of energy
scale corresponds to a shift of $\ga$ by an amount proportional to
$\om$ (\cnf section \ref{sec:s0}).  For energy scale $s_0=Q Q_0$, one
writes a leading-order kernel with the following structure
\begin{equation}
\label{eq:lund}
  \chi_0^\om = 2\psi(1) - \psi\l(\ga+\hom\r)
   - \psi\l(1-\ga+\hom\r),\qquad\qquad s_0= Q Q_0\,,
\end{equation}
originally discussed in \cite{Lund}. Changing energy scale to $s_0 =
Q^2$ corresponds to the shift $\ga\to\ga-\hom$ (\cnf section
\ref{sec:s0}), and we have
\begin{equation}
\label{eq:lundQsqrd}
  \chi_0^\om = 2\psi(1) - \psi\l(\ga\r)
   - \psi\l(1-\ga+\om\r),\qquad\qquad s_0=Q^2\,.
\end{equation}
Remembering that $-\psi(\ga)\simeq 1/\ga$ for small $\ga$, and
iteratively solving for $\om=\asb\chi$ as we did in
section~\ref{sec:s0}, we find an answer which is free of singularities
stronger than $1/\ga$, and so free of spurious double collinear logs.
The procedure can be repeated for energy scale $s_0=Q_0^2$, expanding
around $\ga=1$ and one finds an answer free of spurious double
anti-collinear logs. Expanding \eqref{eq:lund} to order $\asb$ gives
exactly the same triple poles as in \eqref{eq:chi1s0}.

\subsection{The full resummed answer}
\label{sec:fullrsmd}

Let us first see how to correctly include the energy-scale resummation
in the full kernel.  We start with the modified \textsc{LL}
characteristic function, $\chi_0^\om$, eq.~\eqref{eq:lund} which, as
we have just seen, is free of unwanted double (anti) collinear logs
for the (anti) \textsc{DIS} energy scale choice; $\chi_0^\om$
contains \textsc{NLL} corrections,
\begin{equation}
  \label{eq:chi0omNLL}
  \frac{\chi_0}2 \l( -\psi'(\ga) - \psi'(1-\ga)\r),
\end{equation}
which must be subtracted from $\chi_1$ to avoid double counting:
\begin{equation}
  \label{eq:chi1bar}
  {\chit}_1 = \chi_1  - \frac{\chi_0}2 \l( -\psi'(\ga) -
  \psi'(1-\ga)\r). 
\end{equation}
The quantity ${\chit}_1$ still has quadratic and single divergences at
$\ga=0,1$. In analogy with the single divergences in $\chi_0$, these
need to be `shifted' in order to avoid spurious double-collinear logs
at higher orders when changing energy scale. This is accomplished by
subtracting unshifted divergences and replacing them with shifted
divergences:
\begin{multline}
  \label{eq:chi1tilde}
  {\chit}_1^\om = {\tilde \chi}_1 - A_1(0)\psi'(\ga)
  + A_1(\om)\psi'\l(\ga+\hom\r) 
  - (A_1(0) -  b)\psi'(1-\ga) \\
  + (A_1(\om) - b)\psi'\l(1-\ga+\hom\r) + \frac{\pi^2}{6}\l(\chi_0^\om
  - \chi_0\r). 
\end{multline}
Here we have chosen to use $\psi'(\ga)$ and $-\psi(\ga)$ (in $\chi_0$)
as our quadratic and single `divergences to be shifted'. We could
equally well have used $1/\ga^2$ and $1/\ga$ respectively. The
difference in the final result would amount to collinearly suppressed
\textsc{NNLL} terms. The reason for including $A_1(\om)$ in the
shifted poles is discussed shortly.

To resum the splitting-function and running coupling effects, we have
to ensure that $\chi$ has the following structure around $\ga=0$ and
$\ga=1$,
\begin{subequations}
  \label{eq:chil}
\begin{align}
  \chi(\ga,\om) &\simeq  \frac{1 + \om A_1}{\ga+\hom},\qquad\qquad
  &\ga \ll 1\,,\\ 
  \chi(\ga,\om) &\simeq \frac{1 + \om
    (A_1-b)}{1-\ga+\hom},\qquad\qquad &1-\ga   \ll 1\,,
\end{align}
\end{subequations}
where the poles have been shifted compared to eqs.~\eqref{eq:allcoll}
and \eqref{eq:allantcoll} to take into account that they have been
written for energy scale $s_0=Q Q_0$. This can be obtained by writing
\begin{equation}
  \label{eq:chifinal}
  \chi(\ga,\om) = \chi_0^\om + \om \frac{\chit_1^\om}{\chi_0^\om}\,.
\end{equation}
Since $\om =\asb\chi_0^\om + \cO{\asb^2}$ the expansion of $\chi$ to
order $\asb$ is correct. Additionally the ratio
$\chit_1^\om/\chi_0^\om$ contains (shifted) single poles at $\ga=0$
and $\ga=1$ with coefficients $A_1$ and $A_1-b$ respectively, as
required by eqs.~\eqref{eq:chil}. The full $\om$ dependence of $A_1$
is included through the $A_1(\om)$ factors eq.~\eqref{eq:chi1tilde}.

A point to note is that \eqref{eq:chifinal} is no longer an expansion
in $\asb$, but rather in $\om$. For this reason this resummation
technique is known as the $\om$-expansion \cite{CCS}.


\subsection{Results}
\label{sec:res}

\begin{figure}[htbp]
  \begin{center}
    \epsfig{file=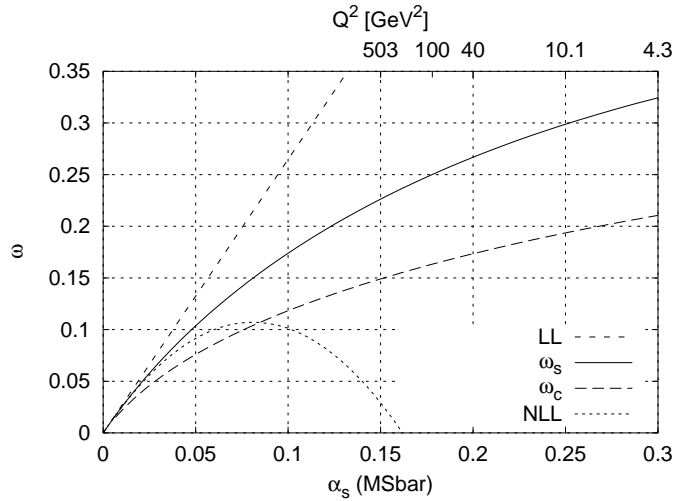,width=0.7\textwidth}
    \caption{Various \textsc{BFKL}  exponents}
    \label{fig:expnts}
  \end{center}
\end{figure}

Figure~\ref{fig:expnts} shows various \textsc{BFKL} exponents as a
function of $\asb$, including the \textsc{LL} and \textsc{NLL} results
for reference. The quantity labelled $\om_s$ is the minimum value of
$\om=\asb\chi(\ga,\om)$, and as such corresponds to the exponent
expected for the gluon Green function at high energies. It is the
power that one expects to observe in $\ga^*\ga^*$ collisions or
forward-jet and Mueller-Navelet jet observables at $ep$ and $pp$
colliders respectively \cite{CCS2}.

Also plotted is a second quantity labelled $\om_c$. This corresponds
to the position of the singularity of the gluon anomalous dimension,
\ie the power growth of small-$x$ splitting functions. Though we have
not really discussed the resummed gluon anomalous dimension, it is
worth noting that $\om_c$ is significantly different from $\om_s$
because it contains additional corrections ${\cal O}(\asb^{5/3})$,
which arise because the effective scale for \textsc{BFKL} evolution in
the anomalous dimension turns out, dynamically, to be considerably
higher than $Q^2$.  Corrections of this form were first noticed in
\cite{HR}.  In general, such corrections are present for quantities
involving an effective cutoff on the lowest accessible transverse
momentum. Another example of such a quantity is the elastic-scattering
cross section.  It should be emphasised therefore that the difference
between $\om_s$ and $\om_c$ is not an uncertainty on the \textsc{BFKL}
exponent, but rather reflects differences between various processes.

The actual uncertainty on the results can be determined by examining
the effect of scheme changes and different approaches to the details
of the resummation (as discussed for example in the previous section,
with regards to the shifting of divergences), as well as a study of
solvable models \cite{CCS2} or other possible higher-order effects
\cite{FRS}. For $\asb\simeq0.2$ it is about $15\%$.

\section{Conclusions and Outlook}
\label{sec:outlook}

In these lectures, we have seen how to deduce many of the properties
of the \textsc{BFKL} pomeron.  The recurrent theme has been the study
of the collinear (and anti-collinear) limit, which gives information
about the structure of divergences of the \textsc{BFKL} characteristic
function at all orders. At \textsc{NLL} order the information thus
obtained is sufficient to reproduce the true \textsc{NLL} corrections
to a high degree of accuracy, \ie the non-collinearly enhanced
\textsc{NLL} corrections are small. Ensuring that the \textsc{BFKL}
kernel correctly reproduces the collinear limit at all orders leads to
stable predictions for the high-energy power growth. The resulting
resummed power is much more compatible with the data than either the
LL or NLL values.

For actual phenomenology, two more ingredients are required. First we
should understand the exponentiation of the characteristic function,
because the running of the coupling complicates the simple approach
that we had at leading order --- it turns out however that these
complications are not too severe \cite{KoMu,Levin,CCS2}.

Secondly we need to know the virtual photon impact factors, \ie the
coupling of a virtual photon (\textsc{DIS} or $\gaga$) to the gluon
chain.  These have still to be worked out at \textsc{NLL}. When
results are eventually available it is likely that a collinear
resummation will again be needed in order to obtain stable
predictions, in analogy with the situation for the characteristic
function.

The overall message is that despite initial fears, the large size of
the \textsc{NLL} corrections \textsc{BFKL} is not an impediment to the
use of \textsc{BFKL} resummation for predicting high-energy phenomena.
But it is necessary to understand the origin of the large corrections,
and include at all orders the physics which causes them.

\section*{Acknowledgements}

Some of the results presented here were obtained in collaboration with
Marcello Ciafaloni and Dimitri Colferai. In writing these lectures I
have benefited also from conversations with Martin Beneke, Carlo Ewerz
and Maneesh Wadhwa.  Finally I would like to thank the organisers of
the School for the welcoming and stimulating environment that they
provided.



\begin{thebibliography}{99}

\bibitem{BFKL} L.N. Lipatov, \sjnp{23}{1976}{338};\\
       E.A. Kuraev, L.N. Lipatov and V.S. Fadin, \jetp{44}{1976}{443};\\
       E.A. Kuraev, L.N. Lipatov and V.S. Fadin, \jetp{45}{1977}{199};\\
       Ya. Balitskii and L.N. Lipatov, \sjnp{28}{1978}{822}.


\bibitem{H1omeff}  H1 Collaboration (Adloff et al.), \npb{470}{1996}{3}
  [\hepex{9603004}]. 


\bibitem{ZEUSF2} ZEUS Collaboration (Breitweg et al.),
  \plb{407}{1997}{402} [\hepex{9706009}].

\bibitem{L3}  L3 Collaboration (M. Acciarri et al.), \plb{453}{1999}{333}.

\bibitem{OPAL} OPAL collaboration, physics note 391, contribution to
  Lepton-Photon '99, Stanford, August 1999.

\bibitem{FJ} ZEUS Collaboration (J. Breitweg et al.), \epjc{6}{1999}{239};\\ 
                    H1 Collaboration (C. Adloff et al.), \npb{538}{1999}{3}.



\bibitem{NLL}
  L.N. Lipatov and V.S. Fadin, \sjnp{50}{1989}{712};\\
  V.S. Fadin, R. Fiore and M.I. Kotsky, \plb{359}{1995}{181};\\ 
  V.S. Fadin, R. Fiore and M.I. Kotsky, \plb{387}{1996}{593}
        [\hepph{9605357}];\\ 
  V.S. Fadin, and L.N. Lipatov, \npb{406}{1993}{259};\\
  V.S. Fadin, R. Fiore and A. Quartarolo, \prd{50}{1994}{5893}
        [\hepth{9405127}];\\
  V.S. Fadin, R. Fiore, and M. I. Kotsky, \plb{389}{1996}{737}
        [\hepph{9608229}];\\
  V.S. Fadin and L.N. Lipatov, \npb{477}{1996}{767} [\hepph{9602287}];\\
  V.S. Fadin, M.I. Kotsky and L.N. Lipatov, \plb{415}{1997}{97};\\
  S. Catani, M. Ciafaloni and F.Hautman, \plb{242}{1990}{97};\\
  S. Catani, M. Ciafaloni and F.Hautman, \npb{366}{1991}{135};\\ 
  V. S. Fadin, R. Fiore, A. Flachi, and M. I. Kotsky, \plb{422}{1998}{287}
  [\hepph{9711427}];\\
  V. Del Duca, \prd{54}{1996}{989};\\
  V. Del Duca, \prd{54}{1996}{4474};\\
  V. Del Duca and C.R. Schmidt, \prd{57}{1998}{4069} [\hepph{9711309}];\\
  V. Del Duca and C.R. Schmidt, \prd{59}{1999}{074004} [\hepph{9810215}].

\bibitem{FL}
  V.S. Fadin and L.N. Lipatov, \plb{429}{1998}{127} [\hepph{9802290}].

\bibitem{CC98}
  M. Ciafaloni and G. Camici, \plb{412}{1997}{396} [\hepph{9707390}];\\
  M. Ciafaloni, \plb{429}{1998}{363} [\hepph{9801322}];\\
  M. Ciafaloni and G. Camici, \plb{430}{1998}{349} [\hepph{9803389}]. 

\bibitem{Ross} D.A. Ross, \plb{431}{1998}{161} [\hepph{9804332}].


\bibitem{Levin} E. Levin, \hepph{9806228}.

\bibitem{GPS} G.P. Salam, \jhep{9807}{1998}{19} [\hepph{9806482}].

\bibitem{CCS}   M. Ciafaloni and D. Colferai, \plb{452}{1999}{372}
  [\hepph{9812366}]; \\
  M. Ciafaloni, D. Colferai and G.P. Salam, \hepph{9905566}. 

\bibitem{DGLAP} V.N. Gribov and L.N. Lipatov, \sjnp{15}{1972}{438};\\
        G. Altarelli and G. Parisi, \npb{126}{1977}{298};\\
        Yu.L. Dokshitzer, \jetp{46}{1977}{641}. 

\bibitem{AbSt} M. Abramowitz and I. Stegun,
        {\it Handbook of Mathematical Functions}, Dover Publications.

\bibitem{DLorig} A. De Rujula, S.L. Glashow, H.D. Politzer,
  S.B. Treiman, F. Wilczek and A. Zee, \prd{10}{1974}{1649}.

\bibitem{GRV} M. Gluck, E. Reya and A. Vogt, \zpc{67}{1995}{433}.

\bibitem{BF} R.D. Ball and S. Forte, \plb{335}{1994}{77}
  [\hepph{9405320}].

\bibitem{dipoleBFKL} A.H. Mueller, \npb{415}{1994}{373}; \\A.H.
  Mueller and B. Patel, \npb{425}{1994}{471}
  [\hepph{9403256};\\
  N.N. Nikolaev and B.G. Zakharov, \jetpl{59}{1994}{6}
  [\hepph{9312268}]; \\ see also the introduction to the dipole
  approach in R.K. Ellis, W.J. Strling and B.R. Webber, {\em QCD and
    Collider Physics}, Cambridge University Press, 1996.

\bibitem{FRbook} J.R. Forshaw and D.A. Ross, {\em Quantum
    Chromodynamics and the Pomeron}, Cambridge University Press, 1997.

\bibitem{BFKLScaling} J. Kwiecinski, \zpc{29}{1985}{561};\\
      J.C. Collins and J. Kwiecinski, \npb{316}{1989}{307}.



\bibitem{CCS2} M. Ciafaloni, D. Colferai and G.P. Salam,
  \jhep{10}{1999}{017} [\hepph{9907409}].  

\bibitem{BFBFKL} R.D. Ball and S. Forte, {\it Proceedings of the DIS 96
             Workshop}, 1996, p.\ 208 [\hepph{9706291}].
  
\bibitem{Maneesh} P. Achard (for the L3 collaboration), talk given at
  Photon 99, Freiburg, Germany, hep-ex/9907016;\\ Maneesh Wadhwa,
  talk given at 19th International Conference on Physics in Collision
  (PIC 99), Ann Arbor, MI, hep-ex/9909001.

\bibitem{CMW}
   S. Catani, G. Marchesini and B.R. Webber,  \npb{349}{1991}{635};\\
  Yu.L.\ Dokshitzer, V.A.\ Khoze and S.I.\ Troyan, \pr{53}{1996}{89}.
  
\bibitem{blumvogt} J. Bl\"umlein and A. Vogt, \prd{57}{1998}{1}
  [\hepph{9707488}];\\ 
 J. Bl\"umlein and A. Vogt, \prd{58}{1998}{014020} [\hepph{9712546}].

\bibitem{Schmidt} C. Schmidt, \prd{60}{1999}{074003} [\hepph{9901397}].

\bibitem{BFKLP} S.J. Brodsky, V.S. Fadin, V.T. Kim, L.N. Lipatov and
  G.B. Pivovarov, \jetpl{70}{1999}{155} [\hepph{9901229}]. 

\bibitem{FRS}  J.R. Forshaw, D.A. Ross and A. Sabio Vera,
  \plb{455}{1999}{273} [\hepph{9903390}].  


\bibitem{Lund}
       B. Andersson, G. Gustafson and J. Samuelsson, \npb{467}{1996}{443}.

\bibitem{HR}  R.E. Hancock and D.A. Ross, \npb{383}{1992}{575};\\
  R.E. Hancock and D.A. Ross, \npb{394}{1993}{200};\\ see also \cite{FRbook}.

\bibitem{KoMu}  Yu.V. Kovchegov and A.H. Mueller, \plb{439}{1998}{428}
        [\hepph{9805208}].


\end{thebibliography}
\end{document}